\documentclass{IEEEtran}
\usepackage{amsfonts}
\usepackage{amsmath}
\usepackage{graphicx}
\usepackage{color}
\usepackage{multicol}
\usepackage{amssymb}

\begin{document}

\title{Closed-Form Expressions for Relay Selection with Secrecy Constraints }

\IEEEoverridecommandlockouts
\author{\IEEEauthorblockN{Xiaojun Sun, Chunming Zhao, Ming
Jiang } }
 \maketitle

\begin{abstract}
An opportunistic relay selection based on instantaneous knowledge of
channels is considered to increase security against eavesdroppers.
The closed-form expressions are derived for the average secrecy
rates and the outage probability when the cooperative networks use
Decode-and-Forward (DF) or Amplify-and-Forward (AF) strategy. These
techniques are demonstrated analytically and with simulation
results.
\end{abstract}


\section{Introduction}
Due to the broadcast nature of transmission medium, wireless
communications are susceptible to eavesdropping. Traditional
security mechanisms mainly rely on cryptographic protocols at higher
layers. In contrast with this paradigm, the physical layer security
strategies exploit the randomness of wireless channels, and
significantly strengthen the security of wireless communications
[1]-[14]. Potential benefits of deriving secure information from
physical layer have been reported in [1].

There has been a growing interest in physical layer security. Wyner
introduced wiretap channel model to evaluate secure transmissions at
the physical layer \cite{IEEEhowto:2}. Csiszar generalized it to
broadcast channels \cite{IEEEhowto:3}. Leung-Yan-Cheong defined the
secrecy capacity as the difference between the main Gaussian channel
capacity and the wiretap Gaussian channel capacity
\cite{IEEEhowto:4}. Wei Kang studied secure communications over a
two-user semi-deterministic broadcast channels \cite{IEEEhowto:5}.
Barros generalized the Gaussian wiretap channel model to wireless
quasi-static fading channel \cite{IEEEhowto:6}-\cite{IEEEhowto:7}.
The secure MIMO systems were studied in
\cite{IEEEhowto:8}-\cite{IEEEhowto:9}. Motivated by emerging
wireless application, relay or cooperative strategies are exploited
to increase security against eavesdroppers
\cite{IEEEhowto:10}-\cite{IEEEhowto:13}. Lai has shown that secure
communications can take place via untrusty relay nodes jamming
eavesdroppers \cite{IEEEhowto:10}. Recently, physical layer secure
protocols based on Decode-and-Forward (DF) or Amplify-and-Forward
(AF) strategy have been proposed in
\cite{IEEEhowto:11}-\cite{IEEEhowto:14} and trusty relay nodes are
employed. To maximize the secrecy capacity, some power allocation
schemes have been presented for DF or AF strategy in
\cite{IEEEhowto:11} and \cite{IEEEhowto:12}, respectively.

This paper investigates relay selection with secrecy constraints in
dual-hop cooperative networks, which use DF or AF strategy. We
select the relay node with the maximal instantaneous secrecy rate to
retransmit the received messages. We assume that the globe channel
state information (CSI) is available [6]-[14] and the number of
relays with successfully decoding is \emph{a prior} [11][13]. Under
this assumption, the closed-form expressions are derived for the
average secrecy rates and the outage probability. A similar work,
which also considered relay selection for secure cooperative
communications, was presented in \cite{IEEEhowto:14}, but it focused
on jamming and outage probability.
\section{Dual-Hop Relay Wiretap Model}
Fig. 1 shows the half-duplex relay system consisting of one source
(S), $N$ relays (R), one destination (D) and one Eve (E). We assume
that the direct links ($S\to D$, $S\to E$) are not available
[11]-[12][14]. Therefore, S transmits confidential information to D
using a trusty R. A third party (Eve) is capable of eavesdropping on
relay's transmissions.
\begin{figure}
\centering
\includegraphics[width=3in, height=1.5in]{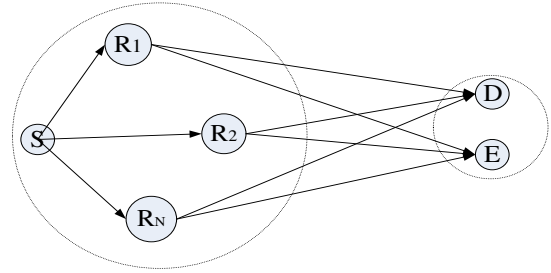}
\vspace{-0.3cm} \caption{Dual-hop relay wiretap channel model.
Define S-R-D link as the main channel and S-R-E link as the wiretap
channel. Eve and destination are located within one cluster while
source is at a faraway location outside the cluster.}
\end{figure}
Suppose that R can access to CSI on wiretap channel and feed back
the CSI to S. This assumption corresponds to the scenario where Eve
is another user interacting with the Time Division Multiple Access
(TDMA) network, thus sending signals that allow R to estimate the
CSI [6][7].

The communication occurs in two hops. In the first hop, S broadcasts
the information to relays. During the second hop, D and Eve observe
the output of the main channel and the wiretap channel from R,
respectively. Denoting $h_{SR,n }$, $h_{RD,n }$ and $h_{RE,n }$ ($n
= 1, \cdots ,N$) as the independent channel gains for S to $n$th R
link ($S \to R_n$), the $n$th R to D link ($R_n \to D$) and to Eve
link ($R_n \to E$), respectively. These channel coefficients are
modeled as zero-mean complex Gaussian random variables. Furthermore,
additive white Gaussian noise (AWGN) is assumed with zero mean and
unit variance. Define the instantaneous signal-to-noise ratio (SNR)
for $S \to R_n$, $R_n \to D$ and $R_n \to E$ as $\gamma _{SR,n }  =
\left| {h_{SR,n } } \right|^2 $, $\gamma _{RD,n }  = \gamma_n \left|
{h_{RD,n } } \right|^2 $ and $\gamma _{RE,n }  = \gamma_n \left|
{h_{RE,n } } \right|^2 $, where $\gamma _n$ is the average SNR. Then
$\gamma _{SR,n }$, $\gamma _{RD,n }$ and $ \gamma_{RE,n } $ are
exponentially distributed random variables with rate parameter $1$,
$\lambda _{m,n }$ and $\lambda _{e,n } $, respectively.

In the second hops, only the R with the largest instantaneous
secrecy rate is selected, which assists S to deliver messages to D
via a DF or AF strategy. The instantaneous secrecy rate about
$n$th-relay link is defined as [6]
\begin{equation}
R_s \left( {Z_n } \right) = \left\{ \begin{gathered}
  \ln Z_n ,\;if\;Z_n  > 1 \hfill \\
  0\;\;\;\;\;,\;if\;Z_n  \leqslant 1 \hfill \\
\end{gathered}  \right.
\end{equation}
where $Z_n=\left( {1 + \gamma _{m,n } } \right)/\left( {1 + \gamma
_{e,n } } \right)$ is the equivalent SNR, $\gamma _{m,n }$ is the
SNR of main channels and $\gamma _{e,n }$ denotes the SNR of wiretap
channels.
\section{Relay Selection with Secrecy Constraints}
This section characterizes the relay selection with secrecy
limitations in terms of average secrecy rate and outage probability
as follows. The equivalent instantaneous SNR at the output of the
relay selection can be expressed as
\begin{equation}
Z_{\max }  = \max \left\{ {Z_1 , \cdots ,Z_N } \right\}
\end{equation}
with cumulative density function (CDF) as
\begin{equation}
F_{\max} \left( {z } \right) = \prod\limits_{n = 1}^N {F_n \left( {z
} \right)}
\end{equation}
where ${F_n \left( {z } \right)}$ is the CDF  of $Z_n$. Then, the
average secrecy rate can be calculated by using (1) and (3)
\begin{equation}
R_s  = \int_0^\infty  {P_r \left( {\ln Z_{\max }  > x} \right)} dx =
\int_0^\infty  {1 - F_{\max } \left( {e^x } \right)} dx
\end{equation}
and the outage probability at a target secrecy rate $R$ can be
written as
\begin{equation}
P_{out} \left( R \right) = P_r \left( {\ln Z_{\max }  \leqslant R}
\right) = F_{\max } \left( {e^R } \right)
\end{equation}
\subsection{Relay Selection Based on DF:S-DF}
The exact expressions for average secrecy rate are calculated in
this subsection. When DF strategy is used, the instantaneous SNR is
$\gamma _{m,n }=\gamma _{RD,n } $ and $\gamma _{e,n }=\gamma _{RE,n
}$. The CDF of $Z_n$ is given by
\begin{equation}
F_n \left( {z } \right) = 1 - \frac{{\lambda _{e,n } \exp \left( { -
\lambda _{m,n }  \left( {z  - 1} \right)} \right)}} {{\lambda _{m,n
} \left( {z  - 1} \right) + \lambda _{m,n }   + \lambda _{e,n }  }}
\end{equation}
\begin{IEEEproof}
We rewrite $Z_n$ as $Z_n  = \left( {1 + \gamma _{RD,n } }
\right)/\left( {1 + \gamma _{RE,n } } \right)=\left( {1 + x }
\right)/\left( {1 + y } \right) $. Since $x$ and $y$ are
independently exponentially distributed random variables, the CDF of
$Z_n$ can be derived as
\begin{equation}
\begin{gathered}
  F_n \left( z \right) = P_r \left( {{Z_n} \leqslant z } \right) \hfill \\
   = \int_0^\infty  {f\left( y \right)} dy\int_0^{yz  + z - 1} {f\left( x \right)} dx \hfill \\
   = 1 - \lambda _{e,n } e^{ - \lambda _{m,n } (z  - 1)} \int_0^\infty  {e^{ - y(\lambda _{m,n } z  + \lambda _{e,n } )} } dy \hfill \\
\end{gathered}
\end{equation}
which yields (6) after some simple manipulations.
\end{IEEEproof}
Using (3), (6) and after applying some algebraic manipulations, we
can express $1 - F_{\max }\left( {e^x } \right)$ as $\sum\limits_i
{\varsigma _i \frac{{\exp \left( { - \beta _i \left( {e^x  - 1}
\right)} \right)}} {{e^x  + \alpha _i }}} $ when channels are
independent but not identically distributed (INID). The integration
in (4) can be rewritten as
\begin{equation}
\begin{gathered}
  R_{sdf}  = \sum\limits_i {\varsigma _i \int_0^\infty  {\frac{{\exp ( - \beta _i \left( {e^x  - 1} \right))}}
{{e^x  + \alpha _i }}dx} } \xrightarrow{{u = e^x  - 1}} \hfill \\
   = \sum\limits_i {\varsigma _i \int_0^\infty  {\frac{{\exp ( - \beta _i u)}}
{{\left( {u + 1 + \alpha _i } \right)\left( {u + 1} \right)}}du} }  \hfill \\
   = \sum\limits_i {\frac{{\varsigma _i }}
{{\alpha _i}}\left[ {F_e \left( {\beta _i } \right) - F_e \left( {\beta _i  + \alpha _i \beta _i } \right)} \right]}  \hfill \\
\end{gathered}
\end{equation}
where $\varsigma _i $, $\beta _i$ and $\alpha _i$ are the
coefficients of the identical equation.
 $F_e \left( x \right) = \exp \left( x \right)E_1 \left( x
\right)$ and $E_1 \left( x \right)$ is the exponential-integral
function [15]. A similar result can be obtained when the channels
are independent identically distributed (IID) by using [15,
Eq.(3.353-2)].
\subsection{Relay Selection Based on AF:S-AF}
Selection AF with the average power scaling (APS) constraint [16]
for secure communications is studied in the following. The
analytical result for AF-APS relay is difficult unless we assume
that the SNR of the $S \to R_n$ links is larger than the SNR of the
$R_n \to D/E$ links [16]. And then, the instantaneous approximate
SNR is $\gamma _{m,n } = \gamma _{SR,n } \gamma _{RD,n } $ and
$\gamma _{e,n } = \gamma _{SR,n } \gamma _{RE,n } $. Let $\mu $
denote $\gamma _{SR,n }$ and use (6) and [15, Eq.(3.324-1)], we can
write the approximate CDF of $Z_n$ as
\begin{equation}
F_n \left( {z } \right) = 1 - \frac{{2  \lambda _{e,n } \sqrt
{\lambda _{m,n }  \left( {z  - 1} \right) } }} {{\lambda _{m,n } z +
\lambda _{e,n } }}{\rm K}_1 \left( {2\sqrt {\lambda _{m,n } \left(
{z - 1} \right)} } \right)
\end{equation}
where $ {\rm K}_1 \left(  \cdot  \right)$ is the 1th-order modified
Bessel function of the seconde kind [15].

Applying (6) and these approximate expressions to (4), we have a
closed-form approximation (10) by using (8) and [15, Eq.(6.565-7)]
\begin{equation}
\begin{gathered}
  R_{saf }  = \sum\limits_i {\frac{{\varsigma _i }}
{{\alpha _i }}\int_0^\infty  {\left[ {F_e \left( {\beta _i /\mu } \right) - F_e \left( {\beta _i \left( {1 + \alpha _i} \right)/\mu } \right)} \right]  e^{ -  \mu } } } d\mu  \hfill \\
   = \sum\limits_i {\frac{{4  \varsigma _i }}
{{\alpha _i }}\left[ {\xi _{1,i} S_{ - 2,1} \left( {\xi _{1,i} } \right) - \xi _{2,i} S_{ - 2,1} \left( {\xi _{2,i} } \right)} \right]}  \hfill \\
\end{gathered}
\end{equation}
where $\xi _{1,i}  = 2\sqrt {  \beta _i} ,\xi _{2,i}  = 2\sqrt {
\beta _i \left( {1 + \alpha _i } \right)} $
 and $S_{a,b} \left(  \cdot  \right)$ is the Lommel
functions [15]. Simulation results show that (10) provides an upper
bound for the average secrecy rate.
\subsection{Optimal Power Allocation for DF:OPA-DF}
For comparison purpose, we just introduced the power allocation
scheme for DF-based protocol [11]. Let us define the $N \times 1$
vectors ${\mathbf{w}} = \left[ {w_1 , \cdots ,w_N } \right]^H $ ,
${\mathbf{h}}_m  = \left[ {h_{RD,1 } , \cdots ,h_{RD,N } } \right]^H
$ and ${\mathbf{h}}_e  = \left[ {h_{RE,1 } , \cdots ,h_{RE,N } }
\right]^H $, the $N \times N$ matrices ${\mathbf{R}}_m =
{\mathbf{h}}_m {\mathbf{h}}_m^H $ and ${\mathbf{R}}_e  =
{\mathbf{h}}_e {\mathbf{h}}_e^H $. For a fix transmit power $\gamma
_0$, the problem of maximizing the secrecy rate $\ln \left[ {(1 +
{\mathbf{w}}^H {\mathbf{R}}_m {\mathbf{w}})/(1 + {\mathbf{w}}^H
{\mathbf{R}}_e {\mathbf{w}})} \right]$ is formulated as
\begin{equation}
\begin{gathered}
  \max \;(1 + {\mathbf{w}}^H {\mathbf{R}}_m {\mathbf{w}})/(1 + {\mathbf{w}}^H {\mathbf{R}}_e {\mathbf{w}}) \hfill \\
  s.t.\;\;{\mathbf{w}}^H {\mathbf{w}} = \gamma _0  \hfill \\
\end{gathered}
\end{equation}
The solution reported in [8][9], is the scaled eigenvector
corresponding to the largest eigenvalue of the symmetric matrix
$\left( {{\mathbf{I}}_N  + \gamma _0 {\mathbf{R}}_m } \right)\left(
{{\mathbf{I}}_N  + \gamma _0 {\mathbf{R}}_e } \right)^{ - 1}$.
\section{Simulation results}
We illustrate the performance of S-DF, S-AF and OPA-DF in this
section. In our test, we assume that all channels are IID Rayleigh
fading and the average wiretap channel SNR is $\gamma _e=10$dB.
Similar conclusions can also be derived when channels are INID and
$\gamma _e$ is other values. We do not plot these curves due to
space limit.
\begin{figure}
\centering
\includegraphics[width=3.0in, height=2.5in]{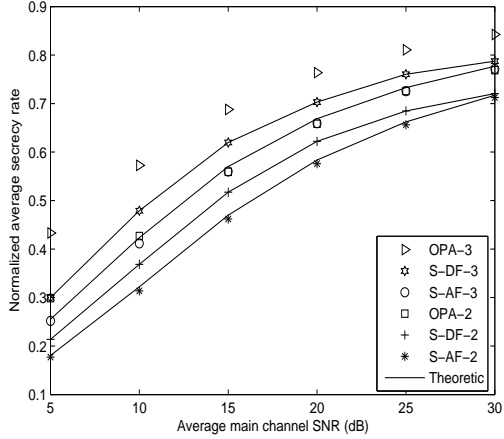}
\vspace{-0.3cm} \caption{The average secrecy rate assuming IID
Rayleigh fading. Normalization is performed with respect to the
capacity of an AWGN channel with the same SNR of main channel.}
\end{figure}
\begin{figure}
\centering
\includegraphics[width=3.0in, height=2.5in]{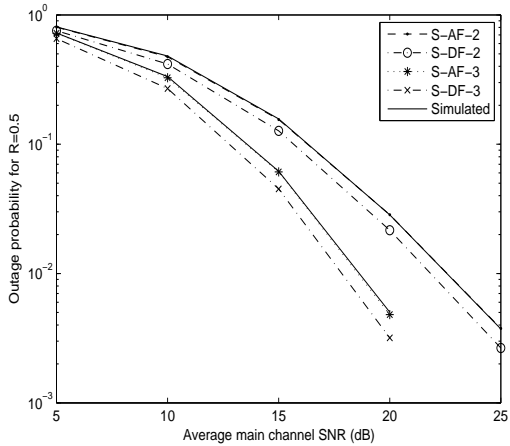}
\vspace{-0.3cm} \caption{Outage probability for a target secrecy
rate 0.5 assuming IID Rayleigh fading.}
\end{figure}

Fig. 2 shows the average secrecy rate of S-DF, S-AF and OPA-DF. The
average secrecy rate of S-DF is larger than that of S-AF. For
comparison purpose, we also plot the curves of OPA-DF. It can be
seen that OPA-DF outperforms S-DF. Compared to selection model, OPA
may have higher implementation complexity, such as, synchronization
of multiple access. The outage probabilities of S-DF and S-AF for a
target secrecy rate 0.5 are depicted in Fig. 3. As expected, S-DF is
also better than S-AF. Taking into account the same number of relay
nodes, we find that the performance of selection AF-APS, which has
the lowest complexity, is the worst. Fig. 2-3 also show that S-AF
may outperform S-DF when the number of AF relay nodes is larger than
that of DF relay nodes.

We can see from Fig. 2-3 that the theoretical results of S-DF are
almost the same as the experimental curves. Simulation results show
that the analytical results of S-AF match exactly with the simulated
curves when the SNR of $S \to R_n$ link is about $16$dB higher than
that of $R_n \to D/E$ links.
\section{Conclusion}
This paper presents the exact mathematical expressions for the
average secrecy rate (ASR) and the outage probability (OP) of
selection DF with secrecy limitations. As a result of computer
simulation, the theoretical results are almost the same as the
experimental curves. The closed-form approximations for the ASR and
OP of selection AF have been derived and match exactly with the
simulated curves when the SNR of $S \to R_n$ link is about $16$dB
higher than that of $R_n \to D/E$ links.


\end{document}